\documentclass{article}

\usepackage{cite}
\usepackage{amsmath,amssymb,amsfonts}
\usepackage{graphicx}
\usepackage{url}
\usepackage[colorlinks=true,linkcolor=blue,citecolor=blue,urlcolor=blue]{hyperref}
\usepackage{cleveref}
\usepackage{geometry}
\usepackage{float}
\usepackage{textcomp}

\usepackage{enumitem}

\usepackage{authblk}

\usepackage{mathtools}
\newcommand{\fxt}[2]{\lceil#1\rfloor^{#2}}
\newtheorem{theorem}{Theorem}
\newtheorem{lemma}{Lemma}
\newtheorem{property}{Property}
\newtheorem{definition}{Definition}

\newtheorem{assumption}{Assumption}

\usepackage[ruled,vlined]{algorithm2e}
\usepackage{algpseudocode}

\newcommand{\sign}{\text{sgn}}

\usepackage{textcomp}
\usepackage{xcolor}
\def\BibTeX{{\rm B\kern-.05em{\sc i\kern-.025em b}\kern-.08em
    T\kern-.1667em\lower.7ex\hbox{E}\kern-.125emX}}
\begin{document}

\title{Robust Fixed-Time Model Reference Adaptive Control}

\author[1]{Chayan Kumar Paul}
\author[2]{Krishanu Nath}
\author[1]{Indra Narayan Kar}
\author[3]{Denis Efimov}
\author[3]{Rosane Ushirobira}

\affil[1]{\small Department of Electrical Engineering, Indian Institute of Technology Delhi, New Delhi 110016, India\\
\texttt{chayanpaul007@gmail.com, ink@ee.iitd.ac.in}}

\affil[2]{\small Department of Electronics and Instrumentation Engineering, NIT Agartala, Tripura 799046, India\\
\texttt{krishanu.eie@faculty.nita.ac.in}}

\affil[3]{\small Inria, Univ.\ Lille, CNRS, UMR 9189--CRIStAL, F-59000 Lille, France\\
\texttt{Denis.Efimov@inria.fr, Rosane.Ushirobira@inria.fr}}

\renewcommand\Authfont{\normalsize}
\renewcommand\Affilfont{\small}
\setlength{\affilsep}{0.6em}

\maketitle

\begin{abstract}
This article proposes a Model Reference Adaptive Control (MRAC) strategy to achieve fixed-time convergence of parameter estimation and tracking errors for unknown linear time-invariant systems, without relying on the persistence of excitation condition. Instead, it employs a less restrictive initial/interval excitation condition on the regressor matrix, enhancing practicality and ease of implementation in real-world scenarios. Our primary contribution is a novel parameter update law within the indirect MRAC framework, ensuring that parameter estimates converge within a fixed time, once the initial/interval excitation condition is met. This approach simplifies the practical requirements for adaptive control while guaranteeing robust performance against parameter uncertainty and external disturbances. Simulation results provide a comparison with the current literature to validate the effectiveness of this approach. Also, implementation on a single-link manipulator is given to illustrate the real-life applicability of this method.
\end{abstract}

\section{Introduction}
Model Reference Adaptive Control (MRAC) \cite{sastry2011adaptive} is a strategy that enables a system to mimic the behavior of a predefined reference model. It can be implemented using two main approaches: direct and indirect. In the direct approach \cite{narendra2003robust}, the adaptive control law is derived directly from the error between the actual system trajectory and the reference model trajectory, leading to a time-varying feedback gain. In contrast, the indirect approach \cite{annaswamy2009model} first estimates the system parameters and then designs the feedback gain based on the estimates. While the direct MRAC approach ensures convergence of the tracking error, it does not provide estimates of the underlying model coefficients. The indirect approach overcomes this limitation by continuously estimating the parameters and updating the control law accordingly, offering a better understanding and adaptability for systems with unknown dynamics.

A key challenge in implementing indirect MRAC is ensuring that the regressor signals satisfy the persistent excitation (PE) condition \cite{bitmead1984persistence,narendra1987persistent}. However, enforcing the PE condition on the regressor signal is often impractical, as it requires the signal to consistently retain sufficient excitation over time -- a requirement that depends on future values, which are challenging to predict or monitor continuously. To address this limitation, concurrent learning (CL) \cite{chowdhary2010concurrent} utilizes stored regressor data that satisfy a rank condition, providing a less restrictive alternative to traditional PE enforcement. While this approach mitigates the strict PE requirement, it is computationally intensive and requires significant memory and processing resources. An alternative strategy, initial (or interval) excitation (IE) \cite{roy2017combined,pan2018composite}, relaxes the PE condition by requiring excitation only within an initial finite time window, eliminating the need for continuous excitation. A sufficiently rich regressor signal fulfills this IE condition almost instantly, and after that, exponential convergence of the parameters is guaranteed. However, since under the IE condition the parameter estimates converge only exponentially, the control input cannot immediately exploit their adaptation, necessitating additional compensatory dynamics to account for the estimation error. Moreover, in the presence of disturbances, the parameter estimation error can only approach a bounded neighborhood around the origin rather than converging exactly to it. The influence of such disturbances on parameter convergence within the IE framework remains largely unexplored.

So far, all strategies have used Lyapunov-based analysis to obtain either asymptotic or exponential convergence results. In 2000, the notion of finite-time convergence \cite{bhat2000finite} gained immense popularity, as it gives stronger convergence-time results than exponential stability, with good robustness. However, in estimation or observer problems, the user usually does not know the actual parameter values or initial error vectors. So, a fixed-time (FxT) \cite{polyakov2011nonlinear} convergence-based approach is usually preferred, in which the estimation error converges within a predefined time independent of the initial condition. 

The concept of fixed-time control has been widely applied to various practical scenarios, including finite-time stabilization of nonlinear systems \cite{hua2016finite}, homogeneous systems \cite{polyakov2016robust,zhang2022fixed}, and consensus tracking \cite{zuo2015nonsingular}. Adaptive backstepping-based FxT control was also proposed for uncertain nonlinear systems \cite{wang2020param}. A general scheme for direct adaptive finite- and fixed-time control is presented in \cite{zimenko2022adaptive}, but no study is conducted on the convergence of the adjusted parameters to their desired values. The application of FxT methods in parameter estimation remains limited. An FxT parameter estimation strategy using dynamic regressor extension and mixing was presented in \cite{korotina2023fixed,wang2020fixed}, proving parameter convergence without strict PE conditions. Yet, it still requires the regressor to be excited throughout the overall estimation period, and no associated control law was developed based on these estimates.

In this article, we present an indirect FxT MRAC scheme for linear systems in the presence of external disturbances, where the regressor matrix satisfies only an IE condition. The stability and convergence properties of the closed-loop system are analyzed using tools from homogeneity theory. Homogeneity, a dilation-based symmetry, is widely exploited in control theory for analysis, control, and observer design \cite{bhat2005geometric,polyakov2016homogeneous,polyakov2019consistent}. In particular, it enables the inference of global stability from local properties and provides an explicit characterization of the convergence rate through the homogeneity degree, thereby offering a systematic and insightful framework for FxT stability analysis.\\
The main contributions in this paper are as follows:
\begin{itemize}
\item A novel fixed-time parameter adaptation law is developed for a class of linear systems under bounded disturbances. The proposed method requires the regressor to satisfy the excitation condition only over an initial finite interval, after which the estimated parameters converge within a fixed time to a small neighborhood of their true values.
\item A fixed-time MRAC framework is formulated by integrating the proposed adaptation law with a model reference controller, ensuring faster transient response and improved robustness against parameter uncertainty against standard MRAC.
\item Homogeneity-based analysis is used to study the closed-loop error dynamics, guaranteeing FxT input-to-state stability (ISS).
\end{itemize}
Additionally, we provide a comparative analysis between the proposed fixed-time parameter convergence law and the exponential convergence update law  \cite{roy2017combined}, demonstrating the efficacy of the proposed approach. The methodology is further validated on a single-link manipulator to assess its practical applicability in real-world settings.

The structure of the paper is as follows. Section \ref{sec:preliminaries} presents an overview of model reference adaptive and FxT controls. Section \ref{sec:fixed_time_identi} introduces the update law for ensuring FxT parameter convergence, while Section \ref{sec:tracking_error} details the control law for trajectory tracking. Section \ref{sec:simultaion} gives comparative analysis on a second-order system through simulations implementation. A conclusion is provided in the last section.

\emph{Notation:} Let $\mathbb{R}_{>0} := \{x \in \mathbb{R} \mid x > 0\}$ denote the set of strictly positive real numbers and $\mathbb{R}_{\geq 0}  := \{x \in \mathbb{R} \mid x \geq 0\}$. For a given norm $\|\cdot \|$ on $\mathbb{R}^n$, $\|A\|$ denotes the induced matrix norm of a matrix $A \in \mathbb{R}^{n \times n}$. The class of continuously differentiable functions is denoted by $C^1$.  The identity matrix of dimension $n$ is denoted by $I_n$. For a matrix $G \in \mathbb{R}^{n \times n}$, its eigenvalues are denoted by $\lambda_i(G)$, $i = 1,\ldots,n$. If $G$ is symmetric, then $\lambda_{\min}(G) := \min_i \lambda_i(G)$ and $\lambda_{\max}(G) := \max_i \lambda_i(G)$. The notation $P > 0$ indicates that a symmetric matrix $P = P^\top \in \mathbb{R}^{n \times n}$ is positive definite. The notation $\Re(\lambda)$ denotes the real part of a complex number $\lambda$.  We denote $\fxt{v}{\alpha}$ $\coloneq \begin{bmatrix}
    |v_1|^\alpha \sign(v_1) &  \dots & |v_n|^\alpha \sign(v_n)
\end{bmatrix}^\top$, for $v = \begin{bmatrix} v_1 & \dots & v_n \end{bmatrix}^\top \in \mathbb{R}^n$.

\section{Preliminaries and Problem Formulation} \label{sec:preliminaries}
\subsection{Preliminaries}
Consider a system with the following dynamics
\begin{equation}
    \dot{y}(t)=F(t,y(t)), \quad y(t_0)=y_0,\; \forall t\geq t_0 \in \mathbb{R}_{\geq 0} \label{eq.sys1}
\end{equation}
with $y(t) \in \mathcal{C}_y \subseteq \mathbb{R}^n$, $\mathcal{C}_y$ an open neighborhood of the origin, and $F: \mathbb{R}_{\geq 0} \times \mathcal{C}_y \to \mathcal{C}_y$ a nonlinear function. Assume that the origin is an equilibrium point of \eqref{eq.sys1}. Refer to \cite{khalil2002nonlinear} for the conventional definitions of stability properties. 

\begin{definition} {\cite{tao2003adaptive}}
    The bounded signal \(y\) is said to be {\em persistently exciting} (PE) if there exist positive scalars $\mu_1$, $\mu_2$, and \(T\in \mathbb{R}_{> 0}\), such that  $\mu_1I_n \leq \displaystyle \int_t^{t+T}y(\tau)y^\top (\tau) \leq \mu_2I_n$,  $\forall t \in \mathbb{R}_{>0}$.
\end{definition}

\begin{definition} \cite{polyakov2011nonlinear}
System \eqref{eq.sys1} is said to be:
    \begin{enumerate}
        \item {\em fixed-time} (FxT) {\em stable}, if it is asymptotically stable, and for any initial condition $y(t_0) \in \mathcal{C}_y$ and $t_0 \in \mathbb{R}_{\geq 0}$, the corresponding solution $y(t)$ of \eqref{eq.sys1} reaches the equilibrium point at some finite-time instant, i.e., $y(t) = 0$, for all $t \geq t_0+ \mathcal{T}_{f}$, where \(\mathcal{T}_{f}\) is called {\em settling time}.
        \item {\em fixed-time} (FxT) {\em attractive} by a bounded set $\mathcal{Y} \subset \mathcal{C}_y$ around the origin, if for any initial condition $y(t_0) \in \mathcal{C}_y$ and $t_0 \in \mathbb{R}_{\geq 0}$, the solution $y(t)$ of \eqref{eq.sys1} reaches the set $\mathcal{Y}$ in some time $t \leq t_0+ \mathcal{T}_{f}$, and it remains inside for all subsequent times, i.e., $y(t) \in \mathcal{Y}$, for all $t \geq t_0+ \mathcal{T}_{f}$.
    \end{enumerate}
\end{definition}

\begin{lemma} \cite{polyakov2011nonlinear}
Let  $V : \mathcal{C}_y \rightarrow \mathbb{R}_{\geq 0}$ be a continuous, positive definite function defined on an open neighborhood of the origin, and $a, b, l_1, l_2 > 0$ such that $0 < l_1 < 1 < l_2$. Assume that any solution $y(t)$ of~(1) satisfies:
\[
\dot{V}(y(t)) \leq -a\,V^{l_1}(y(t)) - b\,V^{l_2}(y(t)).
\]
Then, the system \eqref{eq.sys1} is FxT-stable, and the corresponding settling-time function satisfies \[ \mathcal{T}_{f} \leq \frac{1}{a(1 - l_1)} + \frac{1}{b(l_2 - 1)}.\label{lemma:fxtime} \] 
\end{lemma}

\begin{lemma}
For any vector $y \in \mathbb{R}^n$ and real numbers $0 < r < s$, we have $\|y\|_s \leq \|y\|_r \leq n^{\frac{1}{r} - \frac{1}{s}} \|y\|_s$.
\end{lemma}

This is a direct consequence of the Hölder inequality  \cite{mitrinovic1970analytic}.

Homogeneity is a structural symmetry property of functions, operators, sets, and vector fields with respect to a family of transformations referred to as dilations (see, for example, \cite{bernuau2014homogeneity,andrieu2008homogeneous}). In this work, we consider a one-parameter group of linear dilations $\mathcal{D}: \mathbb{R} \to \mathbb{R}^n$ defined as
\begin{align}
    \mathcal{D}(\tau) := e^{H \tau}
    = \sum_{k=0}^{\infty} \frac{H^k \tau^k}{k!}, \ \forall \tau \in \mathbb{R},
    \label{eq:dilation_definition}
\end{align}
where $H \in \mathbb{R}^{n \times n}$ is a constant matrix so that $-H$ is a Hurwitz matrix and it characterizes the dilation function. The following definitions formalize the notion of homogeneity for dynamical systems.

\begin{definition} \cite{efimov2021finite}

    \begin{enumerate}
    \item A function $h: \mathbb{R}^n \to \mathbb{R}$ is $\mathcal{D}$-{\em homogeneous of degree} $\nu \in \mathbb{R}$ if $ h(\mathcal{D}(\tau)x) = e^{\nu \tau} h(x)$, for all $x \in \mathbb{R}^n$, $\tau \in \mathbb{R}$.

    \item A vector field $f: \mathbb{R}^n \to \mathbb{R}^n$ is $\mathcal{D}$-{\em homogeneous of degree} $\nu \in \mathbb{R}$ if $        f(\mathcal{D}(\tau)x) = e^{\nu \tau}\mathcal{D}(\tau) f(x)$, for all $x \in \mathbb{R}^n$, $\tau \in \mathbb{R}$.

    \item The dynamical system $\dot{x}=f(x)$ is said to be $\mathcal{D}$-{\em homogeneous} if the vector field $f$ is $\mathcal{D}$-homogeneous.
\end{enumerate}

\end{definition}

Let $\mathcal{F}_{\mathcal{D}}(\mathbb{R}^n)$ and $\mathcal{H}_{\mathcal{D}}(\mathbb{R}^n)$ denote the sets of $\mathcal{D}$-homogeneous vector fields and functions which are continuous on $\mathbb{R}^n \setminus \{0\}$, respectively.
The corresponding homogeneity degrees are denoted by $\deg_{\mathcal{F}_{\mathcal{D}}}(\cdot)$ and $\deg_{\mathcal{H}_{\mathcal{D}}}(\cdot)$.

If a $\mathcal{D}$-homogeneous function $h$ is continuously differentiable, then its homogeneity property is inherited by its derivatives. In particular, using Lemma 1 in \cite{zimenko2025homogeneous}, the gradient $\nabla h$ is $\mathcal{D}$-homogeneous of degree $\nu - 1$ in the sense that
\begin{align}
    \nabla h(\mathcal{D}(\tau)x)
    = e^{(\nu - 1)\tau}\mathcal{D}(-\tau)^\top \nabla h(x),
    \quad \forall x \neq 0.
    \label{eq:hom_gradient}
\end{align}

\begin{definition} {\cite{polyakov2016finite}}
    The dilation $\mathcal{D}$ is said to be \emph{strictly monotone} on $\mathbb{R}^n$ if there exists a constant $\kappa \in \mathbb{R}_{>0}$ such that $\|\mathcal{D}(\tau)\| \leq e^{\kappa \tau}$, for all $\tau \leq 0$.
\end{definition}

\begin{lemma} [\cite{polyakov2019sliding}]
    Let $\mathcal{D}$ be a dilation in $\mathbb{R}^n$ with associated matrix $H$. Then:
\begin{itemize}
    \item all eigenvalues $\lambda_i$ of $H$ lie in the
    open right half-plane, that is,
    $\Re(\lambda_i) > 0$, for $i=1,2,\dots,n$;

    \item there exists a symmetric positive definite matrix
    $P = P^\top \in \mathbb{R}^{n \times n}$ such that
    \begin{align}
        P H + H^\top P > 0;
        \label{eq:lyap_ineq}
    \end{align}

    \item the dilation $\mathcal{D}$ is strictly monotone with respect
    to the weighted Euclidean norm
    $\|x\|_P := \sqrt{\langle x,x \rangle_P}$, where $P$ satisfies
    \eqref{eq:lyap_ineq}. In particular,
    \begin{align}
        e^{\kappa_1 \tau} \leq \|\mathcal{D}(\tau)\|_P \leq e^{\kappa_2 \tau},
        &\quad \tau \leq 0, \label{eq:monotone_neg} \\
        e^{\kappa_2 \tau} \leq \|\mathcal{D}(\tau)\|_P \leq e^{\kappa_1 \tau},
        &\quad \tau \geq 0, \label{eq:monotone_pos}
    \end{align}
    where
    \begin{align}
        \kappa_1 &=
        \frac{1}{2}\,
        \lambda_{\max}
        \!\left(
        P^{\frac{1}{2}} H P^{-\frac{1}{2}}
        + P^{-\frac{1}{2}} H^\top P^{\frac{1}{2}}
        \right), \label{eq:kappa1} \\
        \kappa_2 &=
        \frac{1}{2}\,
        \lambda_{\min}
        \!\left(
        P^{\frac{1}{2}} H P^{-\frac{1}{2}}
        + P^{-\frac{1}{2}} H^\top P^{\frac{1}{2}}
        \right).
        \label{eq:kappa2}
    \end{align}
\end{itemize}
\end{lemma}

Homogeneity plays a central role in control theory, as it provides a clear and intuitive way to understand the behavior of nonlinear dynamical systems. In particular, the convergence rate of homogeneous systems can be directly inferred from their homogeneity degree, making homogeneity a practical tool for analyzing finite- and fixed-time stability properties.

\begin{lemma} \cite{zimenko2025homogeneous}
    Consider the $\mathcal{D}$-homogeneous dynamical system:
\[
    \dot{x} = f(x), \qquad f: \mathbb{R}^n \to \mathbb{R}^n,\]
which is asymptotically stable at the origin. Then, the type of convergence is completely characterized by the homogeneity degree $\deg(f) := \deg_{\mathcal{F}_{\mathcal{D}}}(f)$ as follows:
\begin{enumerate}
    \item the system is finite-time stable if $\deg(f) < 0$;

    \item the system is nearly fixed-time stable if $\deg(f) > 0$;

    \item the system is exponentially stable if $\deg(f) = 0$.
\end{enumerate}
\end{lemma}

In addition to convergence rate characterization, homogeneous control systems exhibit strong robustness properties with respect to external disturbances, measurement noise, and time delays (see \cite{BERNUAU20131159,efimov2015weighted,andrieu2008homogeneous,zimenko2023homogeneity}).

\subsection{Problem Formulation}
Consider a continuous linear time-invariant (LTI) system:
\begin{equation}
    \dot{x}=Ax+Bu+d, \label{eq.system}
\end{equation}
where \( x(t) \in \mathcal{C}_x \subseteq  \mathbb{R}^n \) denotes the state vector, and \( u(t) \in \mathbb{R} \) represents the control input. The disturbance \(d(t) \in \mathbb{R}^n\) is bounded by a known scalar, i.e., ess \(\sup_{t \geq 0} \left( |d(t)| \right) < \overline{d}\). The pair \( A \in \mathbb{R}^{n \times n} \) and \( B \in \mathbb{R}^{n} \) is controllable and the matrix \(A\) is unknown. The desired closed-loop response is captured using a reference model, given by:
\begin{equation}
    \dot{x}_m=A_mx_m+Br  ,    \label{eq.reference}
\end{equation}
where \( A_m \in \mathbb{R}^{n \times n} \) is a Hurwitz matrix and \( r(t) \in \mathbb{R} \) is a bounded, continuous reference input. The reference state vector, \( x_m(t) \in  \mathcal{C}_x \subset \mathbb{R}^n \), represents the desired trajectory for the system states in \eqref{eq.system}. The control objective is to ensure that the system state $x$ follows the reference $x_m$.

To ensure tracking in absence of the disturbance, i.e., \(d \equiv 0\), an adaptive control law \cite{narendra2012stable}, incorporating both a linear feedback and feedforward term, can be defined as
\begin{equation}
    u=\mathcal{K}_x^\top  x+r ,   \label{eq.control__input}
\end{equation}
where $\mathcal{K}_x(t) \in \mathbb{R}^{n}$ is a time-varying control gain vector. Substituting \eqref{eq.control__input} in \eqref{eq.system} yields 
\begin{equation}
    \dot{x}=(A+B\mathcal{K}_x^\top  )x+Br. \label{eq.system2}
\end{equation}
To ensure that the system \eqref{eq.system2} behaves as the chosen reference model \eqref{eq.reference}, we need the following assumptions  \cite{tao2003adaptive}.

\begin{assumption} \cite{narendra2012stable} \label{assumption1}
    There exist a constant vector $\overline{\mathcal{K}}_x \in \mathbb{R}^{n}$  such that $A+B\overline{\mathcal{K}}_x^\top=A_m$.
\end{assumption}

\smallskip

Using this assumption, the closed-loop system in \eqref{eq.system2} becomes
\begin{equation}
    \label{eq.closedloop}\dot{x}=A_mx+Br+B\widetilde{\mathcal{K}}_x^\top  x,
\end{equation}
where $\widetilde{\mathcal{K}}_x(t) \triangleq \mathcal{K}_x(t)-\overline{\mathcal{K}}_x$.
Using \eqref{eq.reference} and \eqref{eq.closedloop}, the dynamics of the tracking error $e \triangleq x-x_m$ is obtained as 
\begin{equation}
\dot{e}=A_me+B\widetilde{\mathcal{K}}_x^\top  x. \label{eq.errordynamics}
\end{equation}
The standard direct adaptive update laws \cite{narendra2012stable} for $\mathcal{K}_x$ is: 
\begin{align}
    \dot{\mathcal{K}}_x &=-\Gamma_1 xe^\top  P_1B, \label{eq.paraupd1} 
\end{align}
where $\Gamma_1 \in \mathbb{R}_{>0}$, and $P \in \mathbb{R}^{n \times n}$ is a symmetric positive definite matrix satisfying the Lyapunov equation
\begin{equation}
    A_m^\top  P_1+P_1A_m+Q_1=0,  \label{eq.lyap1}
\end{equation}
for any given symmetric positive definite $Q_1\in\mathbb{R}^{n \times n}.$

The control input \eqref{eq.control__input} along with the parameter update law \eqref{eq.paraupd1} ensures asymptotic tracking, i.e., \(e(t)\to 0\) as \(t\to\infty\). In contrast, our objective is to achieve \emph{fixed-time} convergence for both the parameters estimation and the tracking error.
The estimates of the system matrix, denoted as $\widehat{A}$, can be utilized to indirectly estimate the controller gains. The control gain can be defined as:
\begin{align}
    {\mathcal{K}}_x^\top   &= B^{\dagger}(A_m-\widehat{A}), \label{eq.estimatekx}
\end{align}
where \(B^{\dagger}=(B^\top  B)^{-1}B^\top \) is the pseudo-inverse of $B$.

The convergence of \(\mathcal{K}_x\) to the actual value \(\overline{\mathcal{K}}_x\) typically relies on the PE condition, which is difficult to guarantee in practice. Existing solutions based on IE (interval exciting) may relax the PE condition, but still yield exponential parameter convergence, without a trajectory-independent time bound. Therefore, we seek a parameter update law that ensures that the controller parameters converge to their actual values within a fixed time, independent of the initial conditions. In addition, we aim to develop a parameter update law that guarantees convergence of the tracking error within a prescribed time bound, in the absence of disturbances. In the presence of disturbances, the goal is to achieve fixed-time input-to-state stability (FxT-ISS), meaning that both the parameter estimation error and the tracking error converge to prescribed neighborhoods of the origin within a predetermined time, where the size of the neighborhoods is proportional to the disturbance amplitude.

\section{Fixed-time Identification of Plant Parameters}\label{sec:fixed_time_identi}
The plant dynamics in \eqref{eq.system} can be rewritten as:
\begin{equation}
    \dot{x}=\Phi \Theta+Bu+d,  \label{eq.linparam}
\end{equation}
where $\Phi(x) \in \mathbb{R}^{n\times n^2}$ is the known regressor matrix with
\begin{equation}
    \Phi(x)= I_n \otimes x^\top ,
\end{equation}
and \(\otimes\) represents the matrix Kronecker product. The unknown plant-parameter vector \( \Theta \in \mathbb{R}^{n^2} \), comprising all elements of \( A \), is defined as:
\begin{equation}
    \Theta \triangleq 
        \mathrm{vec}(A^\top ) .
\end{equation}
Inspired by \cite{roy2017combined}, the following first-order filters are designed:
\begin{align}
    \dot{N}&=-kN+\Phi,   \qquad N(0)=0, \label{eq.filter1} \\
    \dot{G}_1&=-kG_1+\dot{x},   \qquad G_1(0)=0, \label{eq.filter2}\\
    \dot{G}_2&=-kG_2+Bu, \qquad G_2(0)=0, \label{eq.filter3}
\end{align}
where $k > 0$ is a scalar gain ensuring filter stability, $N(t) \in \mathbb{R}^{n\times n^2}$ represents a filtered regressor satisfying $\|N(t)\| \leq \overline{N}$, $G_1(t)$, $G_2(t) \in \mathbb{R}^n$ denote the filtered state derivative and input, respectively. The filter variables can be expressed as:
\begin{align}
    N(t)&=e^{-kt}\int_0^t  e^{k \tau}\Phi(\tau) d\tau  \label{eq.filterN},\\
    G_1(t)&=e^{-kt}\int_0^t  e^{k \tau}\dot{x}(\tau) d\tau. \label{eq.filterG}\\
    G_2(t)&=e^{-kt}\int_0^t  e^{k \tau}Bu(\tau) d\tau. \label{eq.filterG2}
\end{align}
From \eqref{eq.linparam}, \eqref{eq.filterN} and \eqref{eq.filterG}, the following can be deduced:
\begin{equation}
    G=G_1-G_2=N\Theta +W  \label{eq.thetarel1}
\end{equation}
where $W(t) \in \mathbb{R}^n$ arises due to the disturbances and can be analytically expressed as
\begin{equation*}
    W(t) \triangleq \int_0^t e^{k(\tau-t)}d(\tau) d\tau.
\end{equation*}
By using \(\|d(t)\|\leq \overline{d}\), one can ensure $W(t)\leq \overline{W}(t) \triangleq \frac{\overline{d}}{k} \left( 1-e^{-kt} \right)$.

To obtain $G_1$, one needs to know $\dot{x}$, which is not available in real time. So, we use integration by parts on  \eqref{eq.filterG}:
    \begin{equation}
        G_1(t)=x(t)-e^{-kt}x(0)-kh(t),  \label{eq.filterG1}
    \end{equation}
    where $h(t)\in \mathbb{R}^n$ is the output of the filter equation:
    \begin{equation}
        \dot{h}= -kh+x,\ \ \ h(0)=0. \label{eq.filterG3}
    \end{equation}
Since \( x \) is measurable, equations \eqref{eq.filterG1}, \eqref{eq.filterG3} and \eqref{eq.filter3} can be employed to compute \( G \) online. Consequently, the filter equations \eqref{eq.filterN} and \eqref{eq.filterG} transform the differential equation \eqref{eq.linparam} into the algebraic equation \eqref{eq.thetarel1}, thereby eliminating the requirement for the direct knowledge of \( \dot{x} \).

A gradient-based FxT parameter estimation law, based on \eqref{eq.thetarel1}, can be formulated to estimate $\Theta$, as proposed in \cite{wang2020fixed}. However, such a law necessitates the restrictive excitation condition on $\Phi$ until the parameter converges, which is practically very difficult to achieve. To overcome this limitation, we use the IE condition, which is imposed through the assumption:

\begin{assumption} \cite{roy2017combined} \label{assumptionIE}
    There exists $T$, $\gamma>0$, such that $\Phi$ is interval exciting (IE) with degree of excitation \( \gamma \) on the interval $[0,T]$, i.e.,
    \begin{equation*}
        \int_0^{T} \Phi^\top  (t)\Phi(t) \geq \gamma I_{n(n+1)}. \label{eq.iecondition}
    \end{equation*}
\end{assumption}

Using Assumption~\ref{assumptionIE}, a less conservative approach can be proposed with the following filters
\begin{align}
    \dot{M} &=-kM+ N^\top N,\quad M(0)=0,\label{eq.filterM} \\
    \dot{H} &=-kH+ N^\top G, \quad H(0)=0, \label{eq.filterH}
\end{align}
where $M(t) \in \mathbb{R}^{n^2 \times n^2}$ is the integrated-filtered regressor and $H(t) \in \mathbb{R}^{n^2}$ is the integrated-filtered output. With the help of \eqref{eq.thetarel1}, one can deduce that
\begin{equation}
    H=M \Theta +W_1,     \label{eq.thetarel2}
\end{equation}
with $W_1(t) \triangleq \int_0^te^{k(\tau-t)}N^\top (\tau)W(\tau) d\tau$.

Since the regressor $N$ is bounded and \(W\) is also bounded, one can obtain that \(\|W_1(t)\|\leq \overline{W}_1\) for some \(\overline{W}_1>0\).

The time-varying matrix $M$ has the following properties:
\begin{property}
    $M(t)$ is a positive semidefinite function of time, i.e., $M(t) \geq 0,\ \ \forall t \geq 0.$
\end{property}
\begin{property} \cite{aranovskiy2022preserving} \label{proper2}
    If the initial excitation condition is satisfied, it can be shown
    \begin{equation}
        \lambda_{\min}\left(M(t)\right) \geq\mu= \gamma \sum_{j=1}^qe^{-jk T},\quad  \forall t\geq qT, q\geq 1. \label{eq.IElowerbound}
    \end{equation}
where $\lambda_{\min} \left( M(t) \right)$ denotes the smallest eigenvalue of $M(t)$.
\end{property}

The parameter estimation law is proposed as follows:
\begin{equation}
    \begin{split}
        \dot{\widehat{\Theta}}=&\kappa N^\top  \left(\lceil G-N\widehat{\Theta}\rfloor^{1-\alpha}+\lceil G-N\widehat{\Theta}\rfloor^{1+\alpha}\right) \\& + \Gamma M^\top  \left(\lceil H-M\widehat{\Theta}\rfloor^{1-\alpha}+\lceil H-M\widehat{\Theta}\rfloor^{1+\alpha}\right)   , \label{eq.updlaw}
    \end{split}
\end{equation}
where $\widehat{\Theta}(t)\in \mathbb{R}^{n(n+1)}$ is an online estimate of the unknown plant-parameter vector $\Theta$. The user-defined variables $\kappa$, $\Gamma >0$ and $\alpha \in (0,1)$ are scalar constants.

The plant parameter estimation error is defined as 
\begin{equation}
    \widetilde{\Theta} \triangleq \widehat{\Theta}-\Theta.    \label{eq.thetaerror}
\end{equation}

Using \eqref{eq.thetarel1}, \eqref{eq.thetarel2}, \eqref{eq.updlaw}, and \eqref{eq.thetaerror}, the dynamics of $\widetilde{\Theta}$ becomes
\begin{equation}
    \begin{split}
            \dot{\widetilde{\Theta}}=-&\kappa N^\top  \left(\fxt{N\widetilde{\Theta}+W}{1+\alpha}+\fxt{N\widetilde{\Theta}+W}{1-\alpha}\right)\\ &-\Gamma M^\top \left(\fxt{M\widetilde{\Theta}+W_1}{1+\alpha}+\fxt{M\widetilde{\Theta}+W_1}{1-\alpha}\right). \label{eq.paraerrordynamics}
    \end{split}
\end{equation}

The proof of Theorem \ref{theorem1} is given in the Appendix.

\smallskip

\begin{theorem} \label{theorem1}
Let Assumption \ref{assumptionIE} be satisfied. The parameter estimation error dynamics in \eqref{eq.paraerrordynamics} is uniformly FxT-ISS stable with respect to the disturbance signals. Moreover, in the absence of disturbances, the convergence time is uniformly bounded by
    \begin{equation}
       \mathcal{T}_{\max}
    = \frac{2}{\alpha}
    \left(\frac{1}{\kappa_1}+\frac{1}{\kappa_2}\right)+qT, \label{eq.settlingtimeexample}
    \end{equation}
   where $\alpha \in (0,1),$ $\kappa_1 = 2^{\frac{2-\alpha}{2}}\mu^{2-\alpha} \Gamma\frac{1-\alpha}{2-\alpha}$, $\kappa_2 = 2^{\frac{2+\alpha}{2}}\mu^{2+\alpha}  p \Gamma(n+1)^{\frac{-\alpha}{2(2+\alpha)}}$, and 
\begin{align*}
p=1-c^{2+\alpha}\frac{1+\alpha}{2+\alpha}
-z^{\frac{2+\alpha}{1+\alpha}}\frac{(1+\alpha)^2}{2+\alpha}>0,
\end{align*}
for some positive $c$, $z>0$.
\end{theorem}

\section{Tracking Error Convergence}\label{sec:tracking_error}

The following theorem proves the fixed-time stabilization of the system \eqref{eq.system} using homogeneity theory \cite{zimenko2025homogeneous}. Its proof is given in the Appendix.

\begin{theorem}\label{theorem21}
Consider the system \eqref{eq.system} under the parameter update law \eqref{eq.updlaw}. Assume that the following conditions hold:  \begin{enumerate}[label=\textbf{(C\arabic*)}, leftmargin=1.2cm]
\item
There exist $Y \in \mathbb{R}^{m \times n}$ and $L \in \mathbb{R}^{n \times n}$ that satisfy
\begin{align*}
    A_m L - L A_m + B Y = A_m, \quad
    L B = 0,
\end{align*}
such that $L - I_n$ is invertible and the matrix $G_d \triangleq \varepsilon I_n + \nu L$ is anti-Hurwitz, for some positive scalars $\varepsilon$  and $\nu$.

\item  Given a positive definite function
\[
\phi \in \mathcal{H}_{\mathcal{D}}(\mathbb{R}^n)
\cap C^1(\mathbb{R}^n \setminus \{0\}, \mathbb{R}_{>0}),
\]
with $\deg_{\mathcal{H}_{\mathcal{D}}}(\phi)=1$, associated with the generator $G_d$, there exists a positive scalar $\beta$ such that 
\begin{align}
    \left\|
    \frac{\partial \phi(\xi)}{\partial \xi}
    \right\|_2
    \, \|\xi\|_2
    \le \beta,
    \label{eq:C3_bound}
\end{align}
for all $\xi \in \mathbb{R}^n$ verifying $\phi(\xi)=1$.
\item For some scalars $\iota > \zeta > 0$,  the system of LMIs
\begin{align*}
\begin{bmatrix}
Q & X(A_m+BK_0)^\top +Y^\top B^\top  \\
* & -\chi I_n
\end{bmatrix}
&\le 0,
\\
    (\zeta - \iota) X + \eta I_n \le 0, 
    \qquad X > 0, \quad\chi>0,
\end{align*}
is feasible for some matrices $X \in \mathbb{R}^{n \times n}$ and $Y \in \mathbb{R}^{1 \times n}$, and some scalars
$\chi$, $\eta \in \mathbb{R}_{>0}$, where $K_0 = Y (L - I_n)^{-1},$
\begin{align*}
Q
&= (A_m + B K_0) X + X (A_m + B K_0)^\top
    + B Y  \nonumber \\ &+ Y^\top B^\top
    + \chi \beta
    \left(\frac12 I_n - G_d\right)
    \left(\frac12 I_n - G_d\right)^\top
    + \iota X .
\end{align*}

\item     The disturbance signal $d$ satisfies
{\footnotesize\begin{align*}
    \left\|
    \mathcal{D}\!\left(-\ln \left( \phi(e))\right)\right) d
    \right\|_2
    \le
    \frac{\psi\,\zeta\,\eta\,\phi(e)^{2\nu}
    \left\|
    X^{-\frac{1}{2}}
    \mathcal{D}\!\left(-\ln \left(\phi(e)\right)\right) e
    \right\|_2^{2}}
    {2\left(1+\|\frac12 I_n-G_d\|_2\right)},
\end{align*}}
for $\psi \in (0,1)$ and for all $e \in \mathbb{R}^n \setminus \{0\}$.
\end{enumerate}
Set the control input of the form
\begin{align}
    u = \begin{cases}
        \mathcal{K}_d^\top x+r,  &t\leq \mathcal{T}_{\max}\\
        \mathcal{K}_x^\top x+r+K_0e+\phi(e)^{\nu+\epsilon}K\mathcal{D}(-\ln(\phi(e))e &t>\mathcal{T}_{\max} \label{eq.controlinput}
    \end{cases}
\end{align}
where  $\mathcal{K}_x$ is given in \eqref{eq.estimatekx} and the adaptive gain $\mathcal{K}_d$ is obtained from \begin{align}
    \dot{\mathcal{K}}_d = \Gamma_d(xe^\top P_1B-\sigma \mathcal{K}_d),  \label{eq.kddot}
\end{align}
with $P_1$ introduced in \eqref{eq.paraupd1}. Then, for $K=YP,P=X^{-1},$ the tracking error $e =x- x_m$ stays bounded in the presence of disturbances. Moreover, for $d \equiv 0$, it reaches the set $\mathcal{C}=\left\{e\in \mathbb{R}^n \mid V(e) \leq \upsilon \right\}$, for any $\upsilon\in \mathbb{R}_{\geq 0}$, within time $\mathcal{T}_{\mathrm{final}}$, where 
\begin{align*}
    V(e) &= e^\top  \mathcal{D}^\top (-\ln \left( \phi(e))\right) P\mathcal{D}(-\ln \left( \phi(e)\right)) e \phi(e),
\end{align*}
 and the final reaching time can be obtained as 
\begin{align}
        \mathcal{T}_{\mathrm{final}}&=\frac{\max\left\{\delta_{\min}^{2\nu}, \delta_{\max}^{2\nu}\right\}}{(1-\psi)\iota\nu} \upsilon^{-\nu}+\mathcal{T}_{\max} \label{eq:finaltime}
\end{align}
where \footnotesize $\delta_{\min}
= \min\limits_{\substack{z \in \mathbb{R}^n \\ h(z)=1}}
\left\| X^{-\frac{1}{2}} z \right\|_2,
\quad \delta_{\max}
= \max\limits_{\substack{z \in \mathbb{R}^n \\ h(z)=1}}
\left\| X^{-\frac{1}{2}} z \right\|_2.$
\end{theorem}

\section{Simulation Results} \label{sec:simultaion}
\subsection{Simulation Results}\label{sec:simonly}
The proposed algorithm is validated in simulation using a second-order LTI system subject to a sinusoidal disturbance
$d(t)=\begin{bmatrix} \tfrac12 \sin(50t) \ \sin(50t) \end{bmatrix}^{\top}$.
Based on \eqref{eq.reference} and \eqref{eq.system}, the system and reference model matrices are selected as follows.
\begin{equation}
    A=\begin{bmatrix}
        0 & 1 \\ -5 & -6
    \end{bmatrix}, B=\begin{bmatrix}
        0 \\ 1
    \end{bmatrix}, A_m=\begin{bmatrix}
        0 & 1 \\ -7 & -10
    \end{bmatrix},\nonumber
\end{equation}
with the reference input $r(t)=5+3\sin(t)+5\sin(2t) U(t-5)$, where $U(t)$ is the unit-step function. The control input $u$ is generated using \eqref{eq.controlinput}, with parameter estimates obtained through \eqref{eq.updlaw}. The initial state vector is chosen as $\begin{bmatrix} 5 & 8 \end{bmatrix}^{\top}$. At $t = 0.5$ seconds, the rank of $M$ is equal to four, and its minimum eigenvalue reaches $\gamma = 0.25$. For the simulation, we set $q = 1$, which gives $\mu = 0.022$.  The design gains for the parameter update law are selected as $\kappa = 25$, $\Gamma = 50$, $\alpha = \frac{2}{3}$, $c=0.8$, $z=0.6$, which results the maximum time of parameter convergence to be $\mathcal{T}_{\max}=4.35 \sec$.

For the controller gains appearing in \eqref{eq.controlinput}, the design parameters are selected as $
\epsilon = 0.5$, $\beta =1.5$, $\zeta =0.1$, $\nu = 0.2$, $v = 0.1$, $K_0 = \begin{bmatrix} 7 & 10 \end{bmatrix}$, and $L   = \begin{bmatrix} 2 & 0 \\ 1 & 0 \end{bmatrix}$.
We choose the function $\phi(e)
= \left(
\sum_{i=1}^{n}
|e_i|^{\rho/(1+(i-n)\nu)}
\right)^{1/\rho}$, as proposed in \cite{zimenko2025homogeneous}, with $\rho = 2$.
By solving the LMIs in \eqref{eq:C3_bound}, the following feasible solution is obtained:
\begin{align*}
\chi = 12.6, \quad \eta = 0.14, \quad \iota = 0.25, \quad
X =
\begin{bmatrix}
14.5 & -6 \\
-6 & 5
\end{bmatrix}.
\end{align*}
Consequently, the resulting controller gain is computed as $
K = \begin{bmatrix}
-1.26 & -2.71
\end{bmatrix}$. For time $t\leq \mathcal{T}_{\max}$, the Lyapunov matrix for the direct adaptive case is given as
\[
P_1 =
\begin{bmatrix}
    0.3025 & -0.5\\
    -0.5   & 1.05
\end{bmatrix} \text{ and } \Gamma_d=10I_2.
\]
Using these values, the theoretical upper bounds on the convergence times are obtained as $\mathcal{T}_{\max} = 4.35 \sec$ and    $\mathcal{T}_{\mathrm{final}} = 29.6\sec$.

\begin{figure*}[t]
    \centering
    \begin{minipage}[t]{0.32\textwidth}
        \centering
        \includegraphics[width=\linewidth,height=2.8cm]{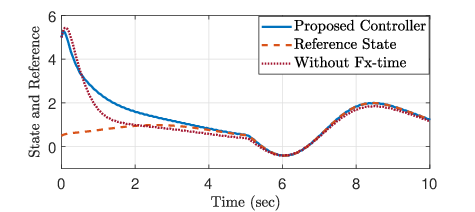}
        \caption{Evolution of the system state \(x_1(t)\) and the reference state \(x_{m_1}(t)\).}
    \label{fig:tracking}
    \end{minipage}
    \hfill
    \begin{minipage}[t]{0.32\textwidth}
        \centering
        \includegraphics[width=\linewidth,height=2.8cm]{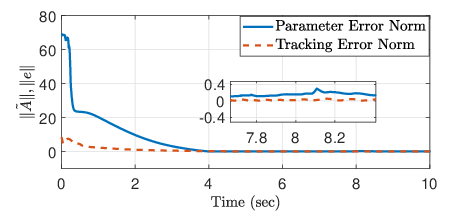}
    \caption{Norm of parameter estimation error and tracking error in presence of disturbance}
    \label{fig:estimation_comparison}
    \end{minipage}
    \hfill
    \begin{minipage}[t]{0.32\textwidth}
        \centering
        \includegraphics[width=\linewidth,height=2.8cm]{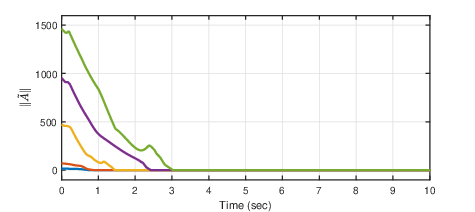}
        \caption{Norm of parameter estimation error for different initial conditions.}
    \label{fig:error_different_initial}
    \end{minipage}

    \vspace{0.5em}

\end{figure*}
As illustrated in Fig.~\ref{fig:tracking}, the system states accurately track the reference trajectory under the proposed control scheme. For comparison, the control strategy presented in \cite{roy2017combined} is also implemented. The resulting plots clearly demonstrate the superior tracking performance of the proposed approach. Fig.~\ref{fig:estimation_comparison} depicts the time evolution of both the parameter estimation error norm and the tracking error norm. Despite the presence of perturbations, the tracking and estimation errors stay bounded, demonstrating the shown input-to-state stability. Moreover, the tracking error is rigorously guaranteed to remain confined within the prescribed set $\mathcal{C}$.\\
To further demonstrate the fixed-time convergence capability, Fig.~\ref{fig:error_different_initial} presents the estimation error trajectories for several different initial conditions. In all cases, the error norm settles within a fixed time, confirming the robustness and consistency of the proposed adaptive framework. This highlights that the method achieves improved tracking performance together with guaranteed fixed-time parameter convergence.

\section{Conclusion}\label{sec:conclusion}
This paper presents a model reference adaptive control (MRAC) algorithm for unknown linear time-invariant systems that eliminates the need for prior knowledge of system matrices or state derivative information. The proposed indirect MRAC framework guarantees fixed-time convergence of parameter estimation and tracking errors in the presence of bounded disturbances. By employing fixed-time convergence, the algorithm ensures faster and more predictable parameter adaptation, resulting in improved tracking performance compared to conventional adaptive control methods. The effectiveness of the proposed method is validated through simulations  implementation, demonstrating its advantages over traditional MRAC techniques. Future work includes extending the approach to output-feedback scenarios and linear time-varying systems for broader applicability.

\bibliographystyle{unsrt}  
\bibliography{references}

@book{narendra2012stable,
  title={Stable adaptive systems},
  author={Narendra, Kumpati S and Annaswamy, Anuradha M},
  year={2012},
  publisher={Courier Corporation}
}

@book{tao2003adaptive,
  title={Adaptive control design and analysis},
  author={Tao, Gang},
  volume={37},
  year={2003},
  publisher={John Wiley \& Sons}
}

@book{mitrinovic1970analytic,
  title={Analytic inequalities},
  author={Mitrinovic, Dragoslav S and Vasic, Petar M},
  volume={1},
  year={1970},
  publisher={Springer}
}

@article{bhat2005geometric,
  title={Geometric homogeneity with applications to finite-time stability},
  author={Bhat, Sanjay P and Bernstein, Dennis S},
  journal={Mathematics of Control, Signals and Systems},
  volume={17},
  number={2},
  pages={101--127},
  year={2005},
  publisher={Springer}
}

@article{efimov2021finite,
  title={Finite-time stability tools for control and estimation},
  author={Efimov, Denis and Polyakov, Andrey and others},
  journal={Foundations and Trends{\textregistered} in Systems and Control},
  volume={9},
  number={2-3},
  pages={171--364},
  year={2021},
  publisher={Now Publishers, Inc.}
}

@article{polyakov2016robust,
  title={Robust stabilization of {MIMO} systems in finite/fixed time},
  author={Polyakov, Andrey and Efimov, Denis and Perruquetti, Wilfrid},
  journal={International Journal of Robust and Nonlinear Control},
  volume={26},
  number={1},
  pages={69--90},
  year={2016},
  publisher={Wiley Online Library}
}

@inproceedings{polyakov2016finite,
  title={On finite-time stabilization of evolution equations: A homogeneous approach},
  author={Polyakov, Andrey and Coron, Jean-Michel and Rosier, Lionel},
  booktitle={2016 IEEE 55th conference on decision and control (CDC)},
  pages={3143--3148},
  year={2016}
}

@article{efimov2015weighted,
  title={Weighted homogeneity for time-delay systems: Finite-time and independent of delay stability},
  author={Efimov, Denis and Polyakov, Andrei and Perruquetti, Wilfrid and Richard, J-P},
  journal={IEEE Transactions on Automatic Control},
  volume={61},
  number={1},
  pages={210--215},
  year={2015},
  publisher={IEEE}
}

@article{zimenko2023homogeneity,
  title={Homogeneity based finite/fixed-time observers for linear {MIMO} systems},
  author={Zimenko, Konstantin and Polyakov, Andrey and Efimov, Denis and Kremlev, Artem},
  journal={International Journal of Robust and Nonlinear Control},
  volume={33},
  number={15},
  pages={8870--8889},
  year={2023},
  publisher={Wiley Online Library}
}

@article{polyakov2016homogeneous,
  title={On homogeneous distributed parameter systems},
  author={Polyakov, Andrey and Efimov, Denis and Fridman, Emilia and Perruquetti, Wilfrid},
  journal={IEEE Transactions on Automatic Control},
  volume={61},
  number={11},
  pages={3657--3662},
  year={2016},
  publisher={IEEE}
}

@article{polyakov2019consistent,
  title={Consistent discretization of finite-time and fixed-time stable systems},
  author={Polyakov, Andrey and Efimov, Denis and Brogliato, Bernard},
  journal={SIAM Journal on Control and Optimization},
  volume={57},
  number={1},
  pages={78--103},
  year={2019},
  publisher={SIAM}
}

@article{polyakov2019sliding,
  title={Sliding mode control design using canonical homogeneous norm},
  author={Polyakov, Andrey},
  journal={International Journal of Robust and Nonlinear Control},
  volume={29},
  number={3},
  pages={682--701},
  year={2019},
  publisher={Wiley Online Library}
}

@article{zhang2022fixed,
  title={Fixed-time and finite-time stability of switched time-delay systems},
  author={Zhang, Junfeng and Efimov, Denis},
  journal={International Journal of Control},
  volume={95},
  number={10},
  pages={2780--2792},
  year={2022},
  publisher={Taylor \& Francis}
}

@article{korotina2023fixed,
  title={Fixed-time parameter estimation via the discrete-time {DREM} method},
  author={Korotina, Marina and Aranovskiy, Stanislav and Ushirobira, Rosane and Efimov, Denis and Wang, Jian},
  journal={IFAC-PapersOnLine},
  volume={56},
  number={2},
  pages={4013--4018},
  year={2023},
  publisher={Elsevier}
}

@article{zimenko2022adaptive,
  title={Adaptive finite-time and fixed-time control design using output stability conditions},
  author={Zimenko, Konstantin and Efimov, Denis and Polyakov, Andrey},
  journal={International Journal of Robust and Nonlinear Control},
  volume={32},
  number={11},
  pages={6361--6378},
  year={2022},
  publisher={Wiley Online Library}
}

@article{pan2018composite,
  title={Composite learning robot control with guaranteed parameter convergence},
  author={Pan, Yongping and Yu, Haoyong},
  journal={Automatica},
  volume={89},
  pages={398--406},
  year={2018},
  publisher={Elsevier}
}

@article{narendra2003robust,
  title={Robust adaptive control in the presence of bounded disturbances},
  author={Narendra, K and Annaswamy, A},
  journal={IEEE Transactions on Automatic Control},
  volume={31},
  number={4},
  pages={306--315},
  year={2003},
  publisher={IEEE}
}

@article{annaswamy2009model,
  title={Model reference adaptive control},
  author={Annaswamy, Anuradha M},
  journal={Control Systems, Robotics and Automation--Volume X: Advanced Control Systems-IV},
  pages={63},
  year={2009},
  publisher={EOLSS Publications}
}

@inproceedings{zimenko2025homogeneous,
  title={Homogeneous control design for linear {MIMO} systems},
  author={Zimenko, K and Polyakov, A and Efimov, D and Ping, X},
  booktitle={2025 European Control Conference (ECC)},
  pages={2770--2774},
  year={2025},
  organization={IEEE}
}

@article{BERNUAU20131159,
title = {Verification of {ISS}, {iISS} and {IOSS} properties applying weighted homogeneity},
journal = {Systems \& Control Letters},
volume = {62},
number = {12},
pages = {1159-1167},
year = {2013},
issn = {0167-6911},
author = {Emmanuel Bernuau and Andrey Polyakov and Denis Efimov and Wilfrid Perruquetti},
  publisher={Elsevier}
}

@book{khalil2002nonlinear,
  title={Nonlinear systems},
  author={Khalil, Hassan K and Grizzle, Jessy W},
  volume={3},
  year={2002},
  publisher={Prentice hall Upper Saddle River, NJ}
}

@article{andrieu2008homogeneous,
  title={Homogeneous approximation, recursive observer design, and output feedback},
  author={Andrieu, Vincent and Praly, Laurent and Astolfi, Alessandro},
  journal={SIAM Journal on control and optimization},
  volume={47},
  number={4},
  pages={1814--1850},
  year={2008},
  publisher={SIAM}
}

@article{bernuau2014homogeneity,
  title={On homogeneity and its application in sliding mode control},
  author={Bernuau, Emmanuel and Efimov, Denis and Perruquetti, Wilfrid and Polyakov, Andrey},
  journal={Journal of the Franklin Institute},
  volume={351},
  number={4},
  pages={1866--1901},
  year={2014},
  publisher={Elsevier}
}

@article{aranovskiy2022preserving,
  title={On preserving-excitation properties of {K}reisselmeier’s regressor extension scheme},
  author={Aranovskiy, Stanislav and Ushirobira, Rosane and Korotina, Marina and Vedyakov, Alexey},
  journal={IEEE Transactions on Automatic Control},
  volume={68},
  number={2},
  pages={1296--1302},
  year={2022},
  publisher={IEEE}
}

@article{bhat2000finite,
  title={Finite-time stability of continuous autonomous systems},
  author={Bhat, Sanjay P and Bernstein, Dennis S},
  journal={SIAM Journal on Control and optimization},
  volume={38},
  number={3},
  pages={751--766},
  year={2000},
  publisher={SIAM}
}

@article{polyakov2011nonlinear,
  title={Nonlinear feedback design for fixed-time stabilization of linear control systems},
  author={Polyakov, Andrey},
  journal={IEEE transactions on Automatic Control},
  volume={57},
  number={8},
  pages={2106--2110},
  year={2011},
  publisher={IEEE}
}

@article{roy2017combined,
  title={Combined {MRAC} for unknown {MIMO} {LTI} systems with parameter convergence},
  author={Roy, Sayan Basu and Bhasin, Shubhendu and Kar, Indra Narayan},
  journal={IEEE Transactions on Automatic Control},
  volume={63},
  number={1},
  pages={283--290},
  year={2017},
  publisher={IEEE}
}

@article{wang2020fixed,
  title={Fixed-time estimation of parameters for non-persistent excitation},
  author={Wang, Jian and Efimov, Denis and Aranovskiy, Stanislav and Bobtsov, Alexey A},
  journal={European Journal of Control},
  volume={55},
  pages={24--32},
  year={2020},
  publisher={Elsevier}
}

@book{sastry2011adaptive,
  title={Adaptive control: stability, convergence and robustness},
  author={Sastry, Shankar and Bodson, Marc},
  year={2011},
  publisher={Courier Corporation}
}

@article{bitmead1984persistence,
  title={Persistence of excitation conditions and the convergence of adaptive schemes},
  author={Bitmead, R},
  journal={IEEE Transactions on Information Theory},
  volume={30},
  number={2},
  pages={183--191},
  year={1984},
  publisher={IEEE}
}

@inproceedings{chowdhary2010concurrent,
  title={Concurrent learning for convergence in adaptive control without persistency of excitation},
  author={Chowdhary, Girish and Johnson, Eric},
  booktitle={49th IEEE Conference on Decision and Control (CDC)},
  pages={3674--3679},
  year={2010}
}

@article{hua2016finite,
  title={Finite/fixed-time stabilization for nonlinear interconnected systems with dead-zone input},
  author={Hua, Changchun and Li, Yafeng and Guan, Xinping},
  journal={IEEE Transactions on Automatic Control},
  volume={62},
  number={5},
  pages={2554--2560},
  year={2016},
  publisher={IEEE}
}

@article{zuo2015nonsingular,
  title={Nonsingular fixed-time consensus tracking for second-order multi-agent networks},
  author={Zuo, Zongyu},
  journal={Automatica},
  volume={54},
  pages={305--309},
  year={2015},
  publisher={Elsevier}
}

@article{wang2020param,
  title={Fixed-time control design for nonlinear uncertain systems via adaptive method},
  author={Wang, Fang and Lai, Guanyu},
  journal={Systems \& Control Letters},
  volume={140},
  pages={104704},
  year={2020},
  publisher={Elsevier}
}

@article{narendra1987persistent,
  title={Persistent excitation in adaptive systems},
  author={Narendra, Kumpati S and Annaswamy, Anuradha M},
  journal={International Journal of Control},
  volume={45},
  number={1},
  pages={127--160},
  year={1987},
  publisher={Taylor \& Francis}
}

\end{document}